\documentclass[aps,prd,reprint,groupedaddress]{revtex4-2}

\usepackage{graphicx}% Include figure files
\usepackage{dcolumn}% Align table columns on decimal point
\usepackage{bm}% bold math

\begin{document}

%\preprint{}

\title{Influence of interstellar environment near the solar system on cosmic ray spectra and dipole anisotropy}


\author{Zhangxi Xue}
\affiliation{School of Physical Science and Technology, Southwest Jiaotong University, Chengdu 610031, People's Republic of China}
\author{Siming Liu}
\affiliation{School of Physical Science and Technology, Southwest Jiaotong University, Chengdu 610031, People's Republic of China}
%\thanks{}


\date{\today}

\begin{abstract}

Properties of interstellar environment near the solar system have been probed by missions like IBEX, Voyager over the last two decades.
Although it has been well recognized that properties of cosmic rays up to the PeV energy can be affected by the local interstellar environment, detailed modeling has not been done. We show that a three component model for the cosmic ray proton and helium spectra from GV to several PV can naturally explain the energy dependence of the dipole anisotropy of cosmic ray fluxes by considering effects of the local interstellar environment on cosmic ray transport, addressing the so-called cosmic ray anisotropy problem. In particular, it is shown that the dipole amplitude and position angle below $\sim 100$ TeV are very sensitive to the velocity of the heliosphere in the local interstellar cloud and the motion of the local interstellar cloud in the local standard of rest. Better measurement of cosmic ray flux anisotropy by experiments like LHAASO and properties of the local interstellar environment by future missions like IMAP will be able to test this model.

\end{abstract}

%\keywords{}

\maketitle

\section{Introduction}

The Earth is located inside the heliosphere about 100 AU from the heliopause, the boundary of the heliosphere \cite{hsieh2022neutral}, and the heliosphere is moving with a speed of $\sim 20$ km s$^{-1}$ at the edge of the Local Interstellar Cloud (LIC) that has a size of a few parseconds (pc) \cite{redfield2000three,linsky2019interface}. A bright ribbon of energetic neutral atom emission with a peak brightness 2 to 3 times greater than the surrounding areas was discovered by the Interstellar Boundary Explorer (IBEX) and is attributed to effects of a large-scale magnetic field around the heliosphere.  
The LIC is moving in the local standard of rest nearly perpendicular to this large-scale magnetic field. The observed properties of cosmic rays will be affected by the above mentioned characteristics of interstellar environment near the Sun \cite{schwadron2014global}.

Galactic cosmic rays (CRs) with energy ranging from GeV to hundreds of TeV may be mainly accelerated by shocks of supernova remnants \cite{2005JPhG...31R..95H,
liu2022origin, 2024SciBu..69.2833C}. Charged CRs, however, change their direction of motion as they propagate in complex magnetic field structure of the Galaxy, losing track of their sources of origin. 
A detailed modeling of the CR transport process is needed to address the origin of Galactic CR.
Thanks to accurate measurements of CR proton and helium spectra by the Alpha Magnetic Spectrometer (AMS) \cite{aguilar2015precision,aguilar2015precisionb}, the Dark Matter Particle Explorer (DAMPE) \cite{dampe2019measurement,alemanno2021measurement}, the CR Energetics And Mass (CREAM) experiment \cite{yoon2017proton}, CR flux anisotropy by a set of ground-based experiments \cite{1997PhRvD..56...23M, bartoli2015argo,bartoli2018galactic,amenomori2005large,amenomori2017northern,aglietta1995study,1996A,aglietta2009evolution,abbasi2010measurement,abbasi2012observation,aartsen2013observation,aartsen2016anisotropy,apel2019search,aab2020cosmic}, and properties of the local interstellar environment by the IBEX \cite{mccomas2009global,schwadron2009comparison,heerikhuisen2009pick,funsten2013circularity,mobius2013analytic,zirnstein2016local} over the last few decades, time is ripe for such an attempt.

Precise measurements of CRs have revealed a spectral hardening at a few hundred GV and a softening near 
$\sim 15$ TV, implying a TeV spectral hump. Recent Large High Altitude Air Shower Observatory (LHAASO) measurements of the all particle spectra and the mean logarithmic mass near the CR spectral knee of a few PeV imply a new ultra-high-energy (with an energy $>0.1$ PeV) component \cite{cao2024measurements}. 
Several models have been proposed to explain these spectral features 
\cite{2012ApJ...752L..13T,  zhang2017anomalous} 
with the TeV spectral hump likely having a local source of origin \cite{2021ApJ...911..151M,
yuan2021nearby, 
%and three components model \cite{
zhang2022three, 
2022ApJ...933...78M}.

Additionally, the observed TeV Galactic CR flux is nearly isotropic with a dipole anisotropy of $\sim 0.1\%$ and a position angle in agreement with the local magnetic field inferred from IBEX observations \cite{schwadron2014global}, which justifies the diffusion approximation for the CR transport \cite{forman1975cosmic}. 
On the other hand, the low value of dipole amplitude implies particular realization of the magnetic field and CR source distribution \cite{mertsch2015solution}. Given the fact that the powerful nearby $\gamma$-ray pulsar Geminga locates close to the direction of the local magnetic field, a three component model was recently proposed to explain the spectra and dipole anisotropy of Galactic CRs \cite{zhang2022three}. The model, however, did not considered the effects of local interstellar environment on the CR properties, leading to a dipole position angles more than $30^\circ$ larger than the observed values. Moreover, the Compton-Getting effect on CR dipole anisotropy caused by motion of the heliosphere in LIC \cite{wood2000heliospheric} has an amplitude comparable to the observed values and a position angle nearly in opposite. This term was either ignored or mistaken in many previous studies \cite{schwadron2014global, yuan2021nearby, zhang2022three}. 
The purpose of this paper is to model the spectra and dipole anisotropy of CRs with full consideration of the effects of local interstellar environment.

\section{Model description}
\label{sec:model}

Our model is similar to that proposed in \citep{zhang2022three} with simplified treatment of the background and Galactic center components.
The background component dominates the flux of GeV CRs and is associated with the Galactic disk, representing the classical view that CRs are accelerated in galactic sources and have been propagating in the Galaxy for tens of millions of years \cite{2022ChPhC..46c0004D}. The TeV spectral hump is attributed to the supernova that gave rise to the Geminga pulsar, and a PeV component from the Galactic center is introduced to account for the amplitude and position angle of dipole anisotropy above $\sim 100$ TeV \cite{zhang2022three} and may be associated with the PeV component inferred from LHAASO measurements of CR spectrum and mean logarithmic mass \cite{cao2024measurements}.

\subsection{Cosmic ray spectra}

The flux intensity of the background component is described with the leaky box approximation \cite{zhang2017anomalous}, i.e.,
\begin{eqnarray}
J_{\rm b}(R)=&&\frac{v\ w\ H_G^2}{4\pi V_GD(R)} \nonumber\\ 
&&\times Q_{\rm 0b}\left(\frac{R}{\rm{GV}}\right)^{-\Gamma}\exp\left(-\frac{R}{R_{\rm cut}}\right), 
\label{eq:bgflux}
\end{eqnarray}
where $R$ is the CR rigidity, 
$v$ is the CR speed, and $w$ is the mean explosion rate of supernovae in the Milky Way galaxy, whose 
thickness and volume are given by $H_G$ and $V_G$, respectively. For typical values of the Galaxy parameters: $H_G=100\;\rm{pc}$, $w=0.03\;\rm{yr^{-1}}$, $V_G= 10\  \rm{kpc^3}$. $\Gamma$ is the spectral index, and the helium spectrum is slightly harder than the proton spectrum with $\Gamma_{\rm He}=\Gamma_{\rm p}-0.077$ \cite{aguilar2015precision,zhang2022three}. The diffusion coefficient is given by 
\begin{equation}
D(R)=D_{\rm 0}\cdot\left(\frac{v}{c}\right)\cdot\left(\frac{R}{10\;\rm{GV}}\right)^{\frac{1}{3}},
\label{eq:diffusioncoefficient}
\end{equation}
where $c$ is the light speed.

We use the force-field approximation to describe the CR modulation by the solar wind, 
so the observed flux near the Earth \cite{gleeson1968solar} is given by
\begin{equation}
J_{\rm b}(R')=\frac{v'R'^{2}}{vR^2}J_{\rm b}(R),
\end{equation}
where 
\begin{equation}
R^2=R'^{2}+2R'\phi\frac{c}{v'}+\phi^2,
\end{equation}
and $\phi$ is the effective potential for the solar modulation.

To fit the TeV spectral bump, we consider an instantaneous injection of particles by a nearby source, presumably the remnant of the supernova that gave rise to the Geminga pulsar with an characteristic age $T_{\rm n}\approx 3.4\times 10^5\;\rm{yr}$. The CRs diffuse along a large scale magnetic field from this source to the solar neighborhood with a distance along the magnetic field $r_{\|}= 250$ pc and perpendicular to the magnetic field $r_{\perp}= 18.5$ pc so that the total energy of injected CRs above 1 GV is about $10^{50}$ ergs for the best fit model \cite{zhang2022three}.  
What's more, considering the fact that the LIC is moving with a velocity ${\bm u}_\mathrm{\bf LSR}$ in the frame of the local standard of rest \cite{redfield2008structure,astra-2-53-2006} and the nearby source is in the upstream of this motion, CR transport from the nearby source will be affected by the local interstellar environment
\cite{schwadron2014global,forman1975cosmic}. The diffusion tensor in a uniform magnetic field can be written as
\begin{equation}
\kappa_{ij}=\kappa_{\perp} \delta_{ij} - \frac{\left(\kappa_{\perp}-\kappa_{\|}\right)B_{i}B_{j}}{B^2},
\label{eq:diffusiontensor}
\end{equation}
where $\delta_{ij}$ is the Kronecker delta tensor, $B_{i}$, and $B_{j}$ are the magnetic field components, and $B$ represents the magnetic field, 
$\kappa_{\|}$, and $\kappa_\perp$
are parallel, and perpendicular 
diffusion coefficients, respectively. In a simple 1D model, CRs from the nearby source 
enter the LIC via the diffusion process \cite{schwadron2014global}, and the flux of CRs from the nearby source \cite{zhang2022three} is given by
\begin{eqnarray}
J_{\rm n}(R)=&&\frac{v}{4\pi} \exp\left(-\frac{u_{\mathrm{LSR}}}{\kappa_{xx}}x\right) \frac{Q_{\rm n}(R)}{(4\pi T_{\rm n})^{3/2}D_{\|}(R)^{1/2}D_{\perp}(R)} \nonumber\\
&&\times\exp\left(-\frac{r_{\|}^2}{4D_{\|}(R)T_{\rm n}}-\frac{r_{\perp}^2}{4D_{\perp}(R)T_{\rm n}}\right),
\label{eq:nearbyflux}
\end{eqnarray}
where the diffusion coefficient $D_{\|}(R)=D_{\rm 0\|}\left(v/c\right)\left(R/10\;\rm{GV}\right)^{1/3}$ and $D_{\rm 0\|}\approx3D_{\rm 0}$,  $D_{\perp}(R)=M_A^4 D_{\|}(R)$ with the magnetic Mach number $M_A=0.1$, which together with $r_\|$ and $r_\perp$ determines contribution of the nearby source to CRs in the solar neighborhood \cite{2008ApJ...673..942Y, zhang2022three}. On large scales, magnetic field wandering is important and the relation between the parallel ($D_\|$) and perpendicular ($D_\perp$) diffusion coefficients are determined by the magnetic Mach number of the background turbulence \cite{2008ApJ...673..942Y, 2019PhRvL.123v1103L, 2022MNRAS.512.2111H}.

Magnetic fields measured by Voyager 1 showed that the local MHD turbulence is generated at a scale of about $10^{14}$ m\cite{lee2020turbulence}. Following \citet{schwadron2014global} with the kinetic theory on small scales, the diffusion coefficient in the direction of ${\bm u}_\mathrm{\bf LSR}$ denoted as $-\bm{\hat{x}}$ is given by
\begin{equation}
\kappa_{xx} =\kappa_{\|}\cos^2\theta+\kappa_\perp\sin^2\theta\,,
\label{eq:kxx}
\end{equation}
where
 $\theta$ is the angle between $\bm{{u}_\mathrm{LSR}}$ and the large-scale magnetic field $\bm{\hat{b}}$  
that has a right ascension of 48.5$^\circ$ and a declination of -21.2$^\circ$ \cite{funsten2013circularity,schwadron2014global}.
The mean distance from the boundary of LIC to the Sun is given by $x$.
$\bm{{u}_\mathrm{LSR}}$ is obtained from the velocity of the LIC with respect to the Sun 
${u}_\mathrm{HC} = 25.4$ km s$^{-1}$ with a right ascension of 
74.9$^\circ$ and a declination of 17.6$^\circ$ \cite{schwadron2015determination} and the solar motion in the local standard of rest. The radial and vertical components $\rm{U_{\odot}}$ and $\rm{W_{\odot}}$ of the solar motion are well determined, but the component $\rm{V_{\odot}}$ in the direction of the Galactic rotation has large uncertainties \cite{dehnen1998local,piskunov2006revisiting,reid2009trigonometric,reid2014trigonometric,xu2018spiral,schonrich2010local} due to its coupling with the Galactic rotation, so we adopt $\rm{U_{\odot}}=11.1$ km s$^{-1}$, $\rm{W_{\odot}}=7.25$ km s$^{-1}$ \cite{schonrich2010local} and treat $\rm{V_{\odot}}$ as a parameter with $\rm{V_{\odot}}=15.5$ km s$^{-1}$ for the best fit. 

We introduce the critical rigidity $R_{\rm cr}=3\;\rm{PV}$ here to characterise the local diffusion coefficients $\kappa_{\|}$ and $\kappa_\perp$. We assume $\kappa_{\|}=k_{0\|}\left(v/c\right)\left(R/10\;\rm{GV}\right)^{1/3}$. $\kappa_{\|}$ and $\kappa_\perp$ are related via $\kappa_\perp=\kappa_{\|}/[1+(\omega \tau)^2]$, where $\omega$ is the gyro-frequency and $\tau=3\kappa_{\|}/v^2$ is the scattering time.
At the critical rigidity $R_{\rm cr}$, the gyro-radius $r_g$ equals to the scattering mean free path $\lambda=\tau v$,
and we get the Bohm diffusion coefficient $k_{\|}=D_{\rm Bohm}=r_gv/3=3.3\times 10^{28}(B/3\;\rm{\mu G})^{-1}\;\rm{cm^{2} \ s^{-1}}$. 
Then we can get $k_{0\|}=4.9\times 10^{26}(B/3\;\rm{\mu G})^{-1}\;\rm{cm^{2}\ s^{-1}}$.
The fact that the diffusion coefficient $k_{0\|}$ in the LIC is smaller than the diffusion coefficient $D_{\rm 0}$ (Table \ref{tab:table1}) outside it may be attributed to enhanced turbulence caused by motion of the LIC. The spectra of the injected particles from the nearby source is given by
\begin{equation}
Q_{\rm n}(R)=Q_{\rm 0n}\left(\frac{R}{\rm{GV}}\right)^{-\Gamma}\exp\left(-\frac{R}{R_{\rm cut}}\right).
\end{equation}
We note that solar modulation is ignored here since the nearby source has few contributions to the low energy cosmic ray fluxes.

The last component of our model is an instantaneous source at the Galactic center with an age $T_{\rm c}= 4\;\rm{Myr}$ and the distance between the Earth and the Galactic center $r_{\rm c}\approx 8.5$ kpc. 
The CR flux 
near the Earth is then given by \cite{gaisser2016cosmic}
\begin{eqnarray}
J_{\rm c}(R)=&&\frac{v}{4\pi} Q_{\rm c}(R) \nonumber\\
&&\times {[4\pi T_{\rm c}\cdot D(R)]^{-3/2}}\exp\left[-\frac{r_{\rm c}^2}{4D(R)T_{\rm c}}\right],
\end{eqnarray}
where the spectra of the injected particles from the Galactic center source is given by
\begin{equation}
Q_{\rm c}(R)=Q_{\rm 0c}\left(\frac{R}{\rm{GV}}\right)^{-\Gamma}\exp\left(-\frac{R}{R_{\rm cutc}}\right).
\end{equation}
The local transport effects are not considered for the background and the Galactic center components since they exist on a much larger scale than the component associated with the nearby source.

\begin{figure}[ht]
\includegraphics[width=8.6cm]{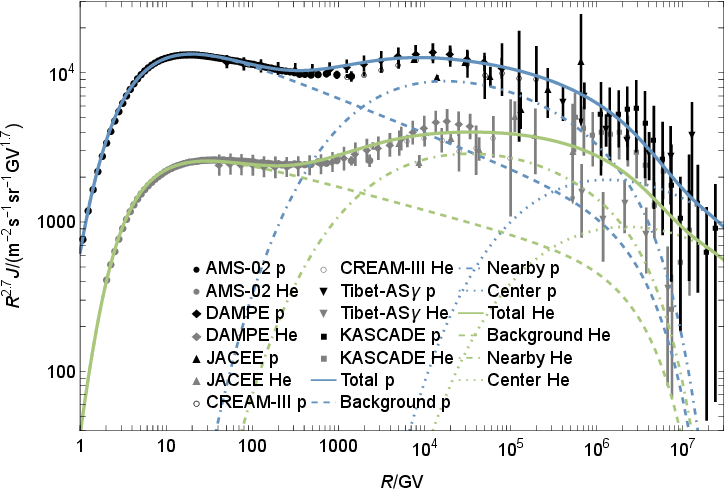}% Here is how to import EPS art
\caption{\label{fig:spectra} The best fit of proton and helium spectra. The data are from AMS \cite{aguilar2015precision, aguilar2015precisionb}, DAMPE \cite{dampe2019measurement, alemanno2021measurement}, JACEE \cite{asakimori1998cosmic}, CREAM \cite{yoon2017proton}, Tibet-AS$\gamma$ results based on SIBYLL + HD \cite{amenomori2006protons}, KASCADE results based on SIBYLL 2.1 \cite{antoni2005kascade}.}
\end{figure}

Proton and helium spectra have been measured precisely and dominate the spectrum of CR, so we mainly consider these two elements.
The best-fitting parameters are given in Table~\ref{tab:table1} and the fitting to the spectra are shown in Fig.~\ref{fig:spectra}, where the blue and green lines represent contributions from proton and helium, respectively. Dashed lines, dot-dashed lines, dotted lines, and solid lines represent the background component, the nearby source, the Galactic center source, and the total spectra, respectively. The fitting of KASCADE spectra based on QGSJet 01 \cite{antoni2005kascade} are shown in the Appendix~\ref{app:spectradata2}, and we also explore the dependence on model parameters in the Appendix~\ref{app:dependence}.

\begin{table}[htbp]
\caption{\label{tab:table1}
The best-fitting parameters of the model.}
\begin{ruledtabular}
\begin{tabular}{ccc}
 Parameters\footnotemark[1] &Best-fitted &Unit\\
\hline
$Q_{\rm 0b,p}$& $4.6\times10^{52}$ &$\rm GV^{-1}$\\ 
$Q_{\rm 0b,He}$& $6.9\times10^{51}$ &$\rm GV^{-1}$\\
$Q_{\rm 0n,p}$& $4.7\times10^{52}$ &$\rm GV^{-1}$\\

$Q_{\rm 0n,He}$& $7.05\times10^{51}$ & $\rm GV^{-1}$ \\
$Q_{\rm 0c,p}$& $5\times10^{57}$ & $\rm GV^{-1}$ \\
$Q_{\rm 0c,He}$& $8\times10^{56}$ & $\rm GV^{-1}$ \\
$R_{\rm cut}$& $4$ & $\rm PV$ \\
$R_{\rm cutc}$& $100$ & $\rm PV$ \\
$\phi$& 0.97 & $\rm GV$ \\
$D_{\rm 0}$ &$2.15\times10^{28}$& \rm{cm$^{2}$ s$^{-1}$}  \\
$\Gamma_{\rm p}$ &$2.6$&   \\
\hline
$\delta_{\rm b}$ &0.0006&   \\
${\rm V}_\odot$ &15.5&   km s$^{-1}$ \\
$x$ &1.6&  \rm pc \\
$u_{\rm HC}$ &25.4&   km s$^{-1}$ \\
$R_{\rm cr}$ & 3 & PV \\
$B$ & 3 & $\mu$G \\
\end{tabular}
\end{ruledtabular}
\footnotetext[1]{Subscripts p and He stand for proton and helium, respectively.}
\end{table}

\subsection{Cosmic ray dipole anisotropy}
As mentioned before, the diffusion tensor is given by equation (\ref{eq:diffusiontensor}), where the antisymmetric part $\kappa_{T}$ has be neglected under the condition that the large-scale magnetic field is uniform \cite{giacalone1999transport}. CR dipole anisotropy is affected by the large-scale magnetic field around the heliosphere in the LIC. According to the standard diffusion theory \cite{schwadron2014global,forman1975cosmic}, the dipole anisotropy of the CR flux distribution caused 
by the nearby source is given by 
\begin{eqnarray}
\bm{{\xi}_{\rm n}}=&&-\frac{3}{v}\left(\frac{\kappa_{\|}}{\kappa_{x x}} u_{\mathrm{LSR}} \cos \theta \bm{\hat{b}}\right.\nonumber\\
&&\left.+\frac{\kappa_{\perp}}{\kappa_{x x}} u_{\mathrm{LSR}} \sin \theta \bm{\hat{e}_{\perp 1,\rm n}}\right),
\end{eqnarray}
where $\kappa_{xx}$ is given by equation (\ref{eq:kxx}).  $\bm{\hat{e}_{\perp 1,\rm n}}=(\bm{\hat{x}}-\cos \theta \bm{\hat{b}})/\sin \theta$ has a right ascension of $53.6^\circ$ and a declination of $68.5^\circ$. 

For the Galactic center source, we can get the dipole anisotropy from the Fick's law \cite{ahlers2017origin}, i.e.,
\begin{equation}
\bm{\delta_{\rm c}}=\frac{3 D(R)
\nabla{J_{\rm c}(R)}}{vJ_{\rm c}(R)} 
= -\frac{3{{r}_{\rm c}}}{2vT_{\rm c}}\bm{\hat{r}_c},
\end{equation}
where $\bm{\hat{r}_c}$ points toward the Galactic center. Here the coordinates are chosen so that the source is located at the distance $r_{\rm c}$ from the Sun. The flow of diffuse CRs is in the same direction as the CR gradient. 
To take into account influence of the large-scale magnetic field on the dipole anisotropy,
we project $\bm{\delta_{\rm c}}$ to the direction of the magnetic field, i.e.,
\begin{eqnarray}
\bm{{\xi}_{\rm c}}=\frac{3{{r}_{\rm c}}}{2vT_{\rm c}}
\cos \theta_{\rm c} \bm{\hat{b}},
\end{eqnarray}
where $\theta_{\rm c}=62.0^{\circ}$ is the angle between the magnetic field $\bm{\hat{b}}$ and the Galactic center dipole anisotropy $\bm{\delta_{\rm c}}$.

For the background component, we assume that the background CRs come from the Galactic disc and propagate perpendicular to it into the LIC, and the dipole anisotropy magnitude is a free parameter $\delta_{\rm b}=0.0006$. The direction of $\bm{\delta_{\rm b}}$ has a Galactic latitude of $90^{\circ}$. Then, we project $\bm{\delta_{\rm b}}$ to the direction of the magnetic field, i.e.,
\begin{eqnarray}
\bm{{\xi}_{\rm b}}=\delta_{\rm b} \cos \theta_{\rm b} \bm{\hat{b}},
\end{eqnarray}
where $\theta_{\rm b}=147.1^{\circ}$ is the angle between the magnetic field $\bm{\hat{b}}$ and the background dipole anisotropy $\bm{\delta_{\rm b}}$. 

Furthermore, the relative motion between the observer and the LIC can lead to the so-called Compton-Getting effect first proposed by Compton and Getting \cite{compton1935apparent}. 
The velocity of the LIC with respect to the Sun is ${u}_\mathrm{HC} = 25.4$ km s$^{-1}$ with a right ascension of 
74.9$^\circ$ and a declination of 17.6$^\circ$ \cite{schwadron2015determination}. The Compton-Getting effect can be written as $3\mathrm{C} \bm{{u}_\mathrm{HC}}/v$, where $\mathrm{C}\simeq1.6$ \cite{schwadron2014global} is one-third of the spectral index of the CR momentum distribution.
Then the overall CR dipole anisotropy reads
\begin{eqnarray}
\bm{{\xi}}(E)=\frac{3}{v}C \bm{{u}_\mathrm{HC}}+
\sum_z\frac{({J_{z\rm n}}\bm{{\xi}_{\rm n}} 
\displaystyle +{J_{z\rm b}}\bm{{\xi}_{\rm b}}+{J_{z\rm c}}\bm{{\xi}_{\rm c}})}{J(E)}\,,
\end{eqnarray}
where $J(E)=\displaystyle \sum_{z}J_{z}=\displaystyle \sum_{z}(J_{z\rm n}+J_{z\rm b}+J_{z\rm c})$ is the overall CR spectrum, $E = [(zeR)^2+m_z^2c^4]^{1/2}$ is the CR energy, $ze$, $m_z$, and $J_z$ stand for the charge, mass, and spectrum of different CR elements, respectively, and the different flux components need to be converted into fluxes with respect to $E$.

\begin{figure}[htbp]
\centering
\begin{minipage}[h]{0.9\columnwidth} 
  \includegraphics[width=\linewidth]{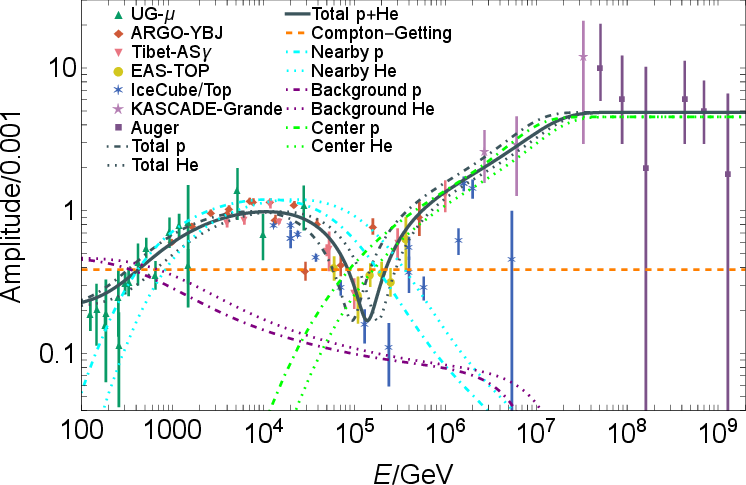}
  %\caption{Top Image}
  %\label{fig:top}
\end{minipage}\par 
\vspace{5mm}
\begin{minipage}[h]{0.9\columnwidth} 
  \includegraphics[width=\linewidth]{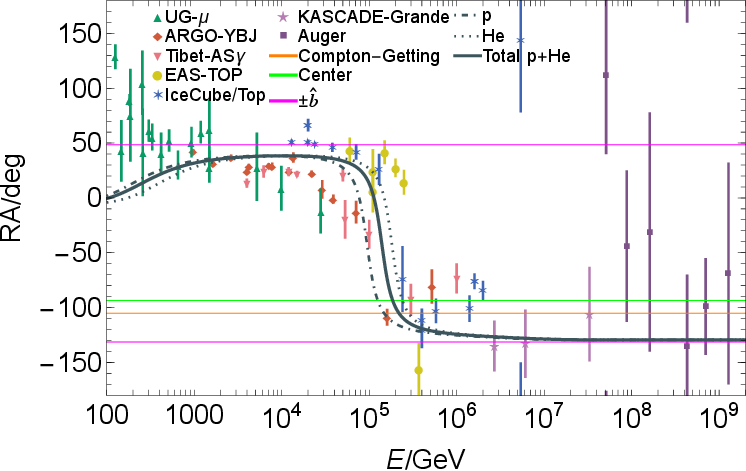}
  %\caption{Bottom Image}
  %\label{fig:bottom}
\end{minipage}
\caption{Upper: The best fit of dipole anisotropy amplitude and contributions of each component to dipole anisotropy amplitude. Bottom: The best fit of dipole anisotropy position angle. The position angles of the Galactic center, Compton-Getting due to motion of the Sun in the LIC, and the magnetic field are also indicated. The data are from UG-$\mu$ \cite{1997PhRvD..56...23M, amenomori2005large}, ARGO-YBJ \cite{bartoli2015argo, bartoli2018galactic}, Tibet-AS$\gamma$ \cite{amenomori2005large, amenomori2017northern}, EAS-TOP \cite{aglietta1995study, 1996A, aglietta2009evolution}, IceCube/Top \cite{abbasi2010measurement, abbasi2012observation, aartsen2013observation, aartsen2016anisotropy}, KASCADE-Grande \cite{apel2019search}, Auger \cite{aab2020cosmic}.}
\label{fig:diploeanisotropy}
\end{figure}
\begin{figure}[htbp]
\includegraphics[width=8cm]{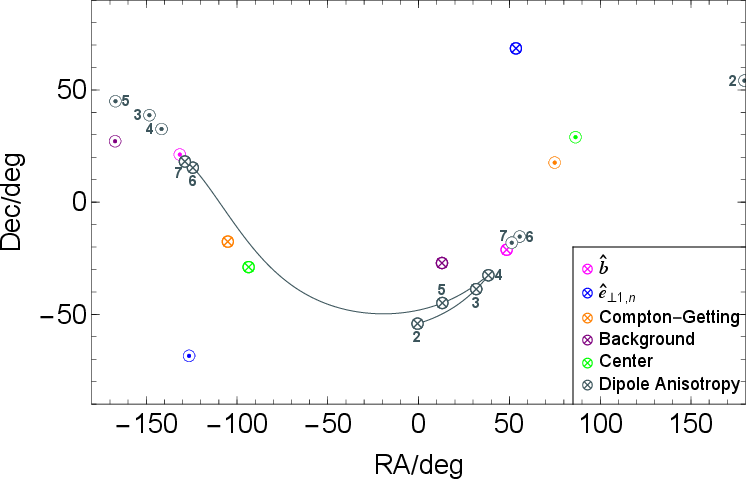}% Here is how to import EPS art
\caption{\label{fig:skymap} Sky map in equatorial coordinates. The circle-cross signs represent the vector directions and the circle-dot signs stand for the reverse. The numbers indicate $\log_{10}\left(E/\rm{GeV}\right)$.}
\end{figure}

Due to limitations of ground-based observatories, we can only get the overall CR dipole anisotropy aligning with the equatorial plane \cite{ahlers2016deciphering}, so we project $\bm{{\xi}}$ onto the equatorial plane when fitting the data. The amplitude and phase fitting results of the dipole anisotropy are shown in Fig.~\ref{fig:diploeanisotropy}, where the solid lines are the overall amplitude and phase of dipole anisotropy of proton and helium, the dot-dashed and dotted lines represent proton and helium, respectively. We also show the contribution of each component in different colors. Fig.~\ref{fig:skymap} shows directions of different components and the dependence of the total dipole anisotropy on energy in the sky map.

In the sub-TeV energy range, the CR flux is dominated by the background component (Fig. \ref{fig:spectra}). Its anisotropy is partially canceled out by the Compton-Getting effect due to their nearly opposite position angles (Fig. \ref{fig:skymap}). Entering the TeV range, the nearby source starts to play the dominant role. In the Appendix~\ref{app:dependence}, we show that the CR spectra and dipole anisotropy are very sensitive to properties of the interstellar environment nearby, which is primarily due to the fact that the LIC is moving in the local standard of rest in a direction nearly perpendicular to the direction of the magnetic field \cite{schwadron2014global}. \citet{mertsch2015solution} showed that the dipole anisotropy can be strongly suppressed if the CR gradient is in perpendicular to the magnetic field. For contributions from the nearby source, the CR gradient is aligned with the motion of the LIC in the local standard of rest, addressing the low amplitude of CR dipole anisotropy. Interestingly, within uncertainties of measured properties of the local interstellar medium, both the spectra and dipole anisotropy below 100 TeV are reproduced. The CR measurements therefore may be used to probe properties of the local interstellar environment.

Above $\sim100$ TeV, the dipole anisotropy shifts toward the Galactic center. Although the Galactic center component starts to dominate the CR fluxes above 1 PeV (Fig.\ref{fig:spectra}), its contribution to the dipole anisotropy becomes important near 100 TeV. Moreover, the nearby source is located in the opposite direction of the Galactic center, a minimum of the dipole amplitude is realized slightly above 100 TeV (Fig.\ref{fig:diploeanisotropy}).

\section{Effects of magnetic field wandering in the local interstellar environment}

In the previous section, we assume that magnetic field wandering is important on large scales and use the kinetic theory (without considering the effects of magnetic field wandering) on small scales. 
DAMPE reported that the proton spectral index is about 2.85 beyond a spectral softening near 13.6 TeV \cite{dampe2019measurement}. If one uses kinetic theory on large scales, then $\kappa_{\|} = \tau v^2 / 3\propto R^{1/3}$, $\kappa_\perp=\kappa_{\|}/[1+(\omega \tau)^2] \propto R^{5/3}$, the proton spectrum will be softer than the injection spectrum by a factor of $\kappa_{\|}^{1/2}\kappa_\perp \propto R^{11/6}$. To fit the observed spectrum, the spectral index of injection needs to be about 1.02, which is too hard to be explained by shocks of supernova remnants. Therefore, magnetic field wandering effects must be considered on large scales.

\begin{figure}[htbp]
\centering
\begin{minipage}[h]{0.9\columnwidth} 
  \includegraphics[width=\linewidth]{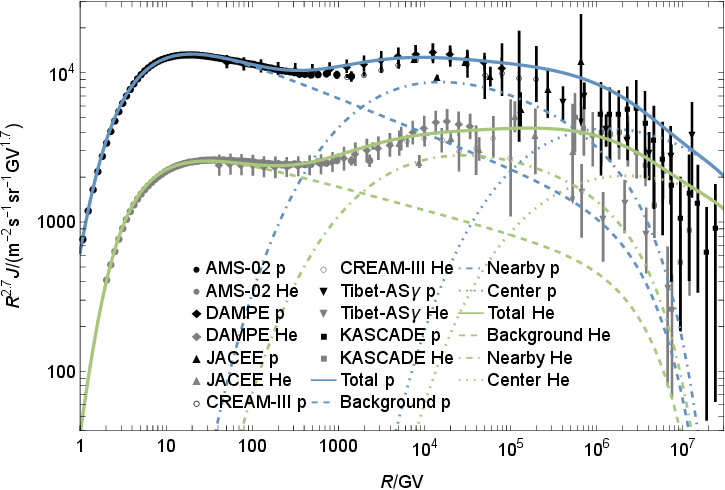}
\end{minipage}\par 
\vspace{5mm}
\begin{minipage}[h]{0.9\columnwidth} 
  \includegraphics[width=\linewidth]{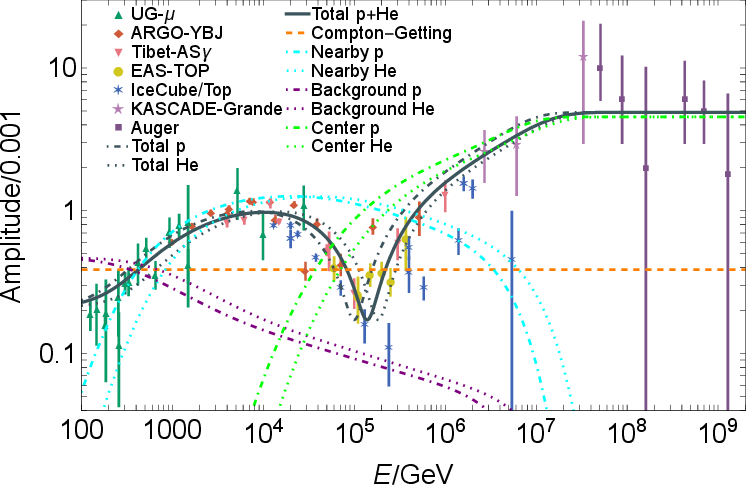}
\end{minipage}\par 
\vspace{5mm}
\begin{minipage}[h]{0.9\columnwidth} 
  \includegraphics[width=\linewidth]{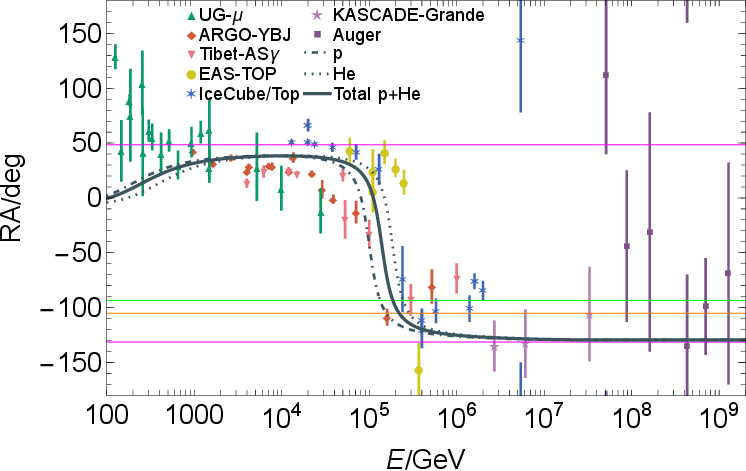}
  %\caption{Bottom Image}
  %\label{fig:bottom}
\end{minipage}
\caption{The fitting results considering magnetic field wandering on both small and large scales. Upper: The best fit of proton and helium spectra. Middle: The best fit of dipole anisotropy amplitude. Bottom: The best fit of dipole anisotropy position angle. The data used here is the same as that in Fig.~\ref{fig:spectra} and Fig.~\ref{fig:diploeanisotropy}.}
\label{fig:DD}
\end{figure}

Although the previous section shows that kinetic theory on small scales can explain the observed properties of CRs,
magnetic field wandering can be important at small scales as well. Then following a similar procedure, the flux of CRs from the nearby source can be written by
\begin{eqnarray}
J_{\rm n}(R)=&&\frac{v}{4\pi} \exp\left(-\frac{u_{\mathrm{LSR}}}{D_{xx}}x\right) \frac{Q_{\rm n}(R)}{(4\pi T_{\rm n})^{3/2}D_{\|}(R)^{1/2}D_{\perp}(R)} \nonumber\\
&&\times\exp\left(-\frac{r_{\|}^2}{4D_{\|}(R)T_{\rm n}}-\frac{r_{\perp}^2}{4D_{\perp}(R)T_{\rm n}}\right),
\label{eq:nearbyfluxDD}
\end{eqnarray}
where $D_{xx} =D_{\|}\cos^2\theta+D_\perp\sin^2\theta$, and the dipole anisotropy contributed by the nearby source is given by
\begin{eqnarray}
\bm{{\xi}_{\rm n}}=&&-\frac{3}{v}\left(\frac{D_{\|}}{D_{x x}} u_{\mathrm{LSR}} \cos \theta \bm{\hat{b}}\right.\nonumber\\
&&\left.+\frac{D_{\perp}}{D_{x x}} u_{\mathrm{LSR}} \sin \theta \bm{\hat{e}_{\perp 1,\rm n}}\right).
\end{eqnarray}
To fit the observations, we use the same parameters as in the previous section except for $r_{\perp}= 20$ pc, $Q_{\rm 0c,p}=1.1\times10^{58}\;\rm GV^{-1}$, and $Q_{\rm 0c,He}=1.76\times10^{57}\;\rm GV^{-1}$. 
Due to more contribution from the nearby source toward higher energies, contributions from the Galactic center component needs to be enhanced to fit the CR dipole anisotropy.
This modeling has fewer parameters and the fitting results are shown in Fig.~\ref{fig:DD}.

\section{Conclusion}

The dipole anisotropy of CR flux shows complicated energy dependent behavior especially in the TeV range, concurrent with a spectral bump. These results strongly suggest contributions from a CR source near the solar system. Propagation of CRs from this nearby source to the Earth will be affected to properties of the interstellar medium in the solar neighborhood. In this work, based on the standard kinetic theory for CR diffusion \cite{giacalone1999transport} and/or considering the effect of magnetic field wandering \citep{2008ApJ...673..942Y}, we use a three-component model for the Galactic CR spectra to show that the dipole anisotropy of CR flux can be reproduced with measured properties of the local interstellar environment. 
Although both diffusion models with or without considering the magnetic field wandering effect at the solar neighborhood can explain relevant observations, magnetic field wandering must be considered in modeling particle transport from the nearby source to the solar neighborhood.
Moreover, since the LIC is moving nearly perpendicular to the magnetic field, the CR dipole anisotropy is very sensitive to these properties. Future observations of the local interstellar medium with the IMAP and CRs with the LHAASO, Auger, IceCube etc will be able to test this model. 

Given huge uncertainties of the proton and helium spectra in the PeV energy range \cite{antoni2005kascade}, no attempt has been made to reproduce the recent precise measurements of CR spectrum and mean logarithmic mass by LHAASO \cite{cao2024measurements}. Heavier elements need to be considered in future studies.

\begin{acknowledgments}
This work is supported by the National Natural Science Foundation of China under the grant No. 12375103.    
\end{acknowledgments}

\appendix

\section{Dependence of the fitting results on model parameters} 
\label{app:dependence}
\subsection{Dependence on the magnitude of the background dipole anisotropy $\delta_b$}
%\begin{widetext}
\begin{figure}[h!]
%\centering
%\begin{minipage}{0.32\textwidth}
%  \centering
  \includegraphics[width=0.32\linewidth]{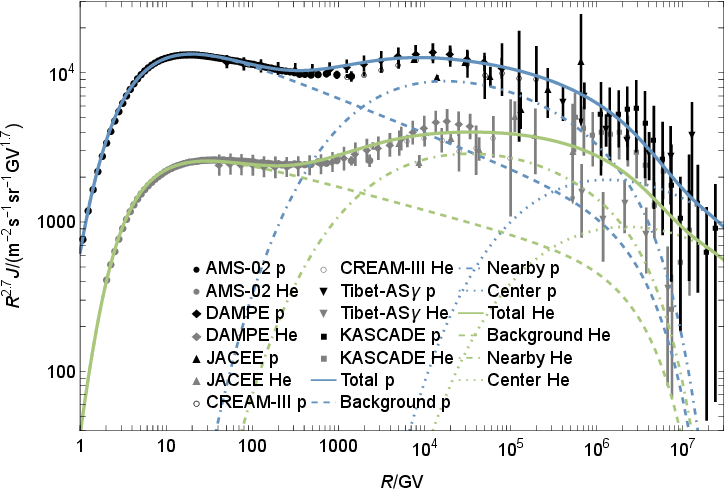}
%\end{minipage}
%\begin{minipage}{0.32\textwidth}
%  \centering
  \includegraphics[width=0.32\linewidth]{spectra.eps}
%\end{minipage}
%\begin{minipage}{0.32\textwidth}
%  \centering
  \includegraphics[width=0.32\linewidth]{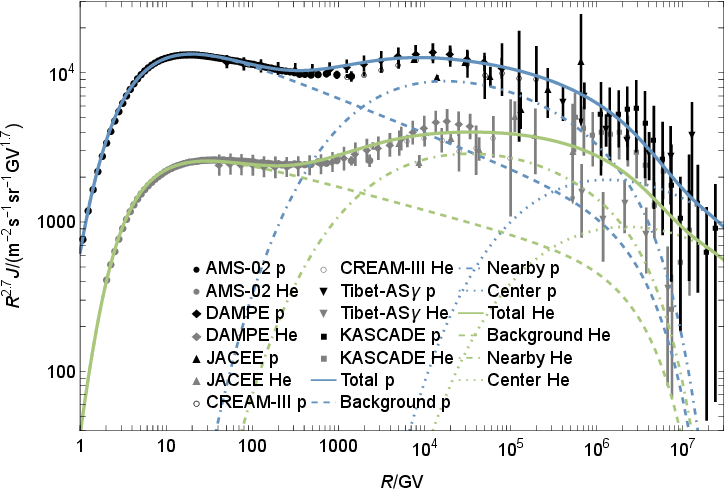}
%\end{minipage}
%\begin{minipage}{0.32\textwidth}
%  \centering
  \includegraphics[width=0.32\linewidth]{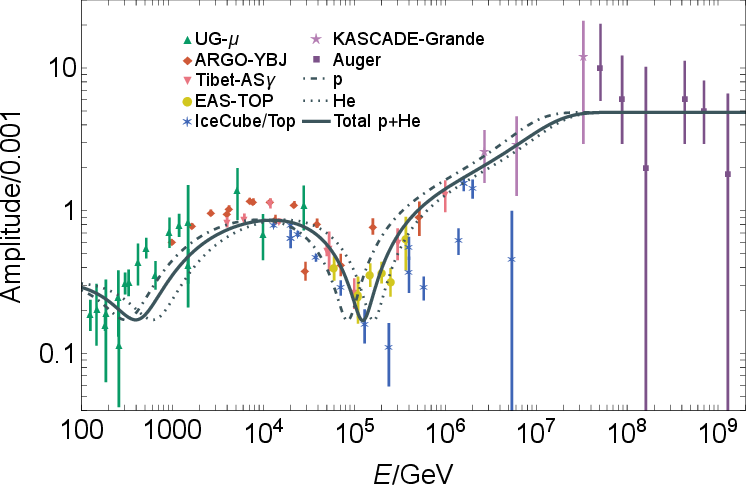}
%\end{minipage}
%\begin{minipage}{0.32\textwidth}
%  \centering
  \includegraphics[width=0.32\linewidth]{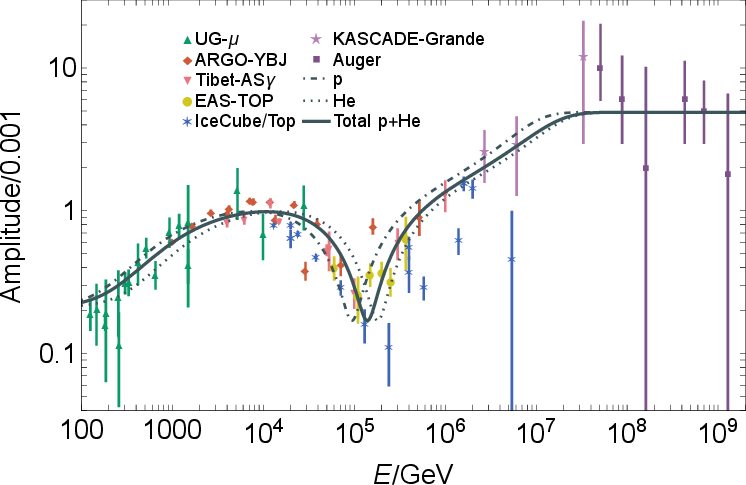}
%\end{minipage}
%\begin{minipage}{0.32\textwidth}
%  \centering
  \includegraphics[width=0.32\linewidth]{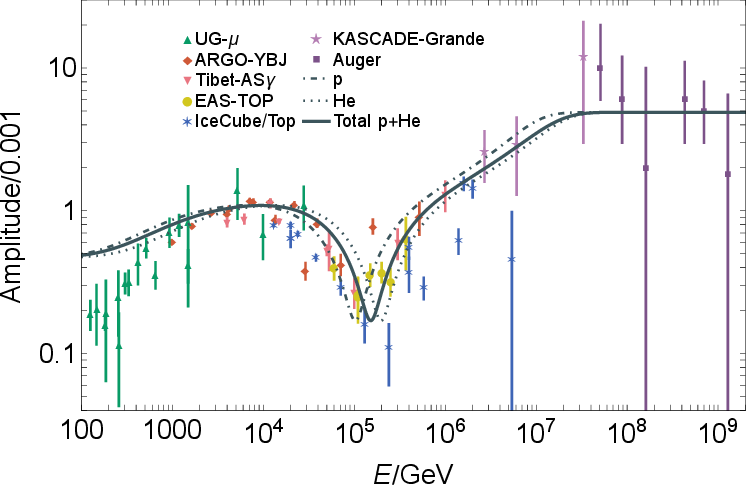}
%\end{minipage}

%\begin{minipage}{0.32\textwidth}
%  \centering
  \includegraphics[width=0.32\linewidth]{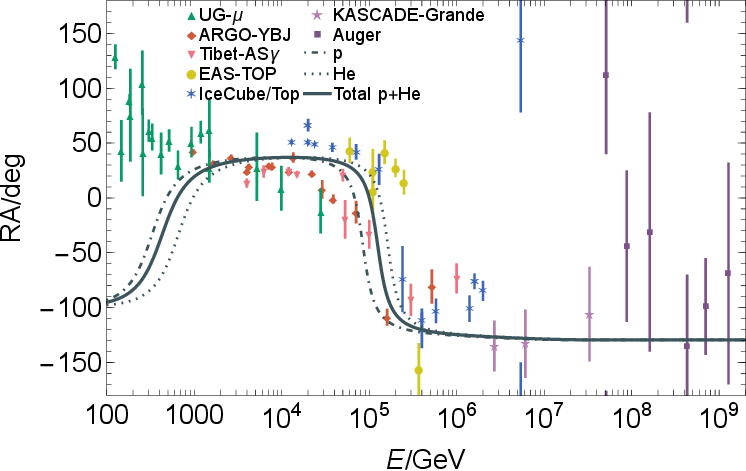}
%\end{minipage}
%\begin{minipage}{0.32\textwidth}
%  \centering
  \includegraphics[width=0.32\linewidth]{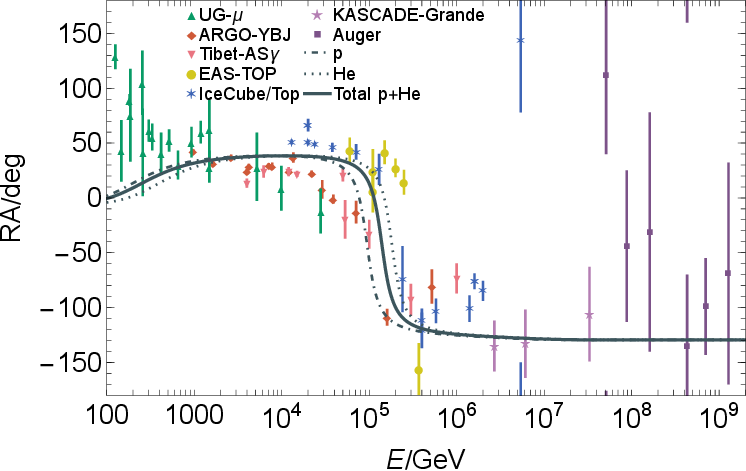}
%\end{minipage}
%\begin{minipage}{0.32\textwidth}
%  \centering
  \includegraphics[width=0.32\linewidth]{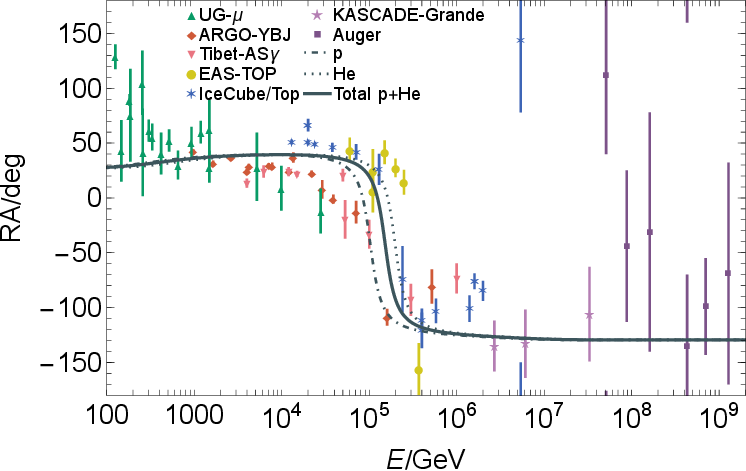}
%\end{minipage}
\caption{Influences of $\delta_b$ on fitting results. Left column: Fitting results with $\delta_b = 0.0001$. Middle column: The best-fitting results with $\delta_b = 0.0006$. Right column: Fitting results with $\delta_b = 0.001$.}
\label{fig:delta}
\end{figure}
%\end{widetext}

In our model, $\delta_b$ is a free parameter due to our approximate treatment of the background component. Figure \ref{fig:delta} shows that it can be well constrained by observations and mostly affect the overall anisotropy in the sub-TeV energy range where the background component dominates. Such a component plays a crucial role in canceling out the Compton-Getting effect at low energies, leading to the observed low amplitudes and position angles.

\subsection{Dependence on solar motion in the direction of Galactic rotation $\rm{V_{\odot}}$}
\begin{figure}[h!]
%\centering
%\begin{minipage}{0.32\textwidth}
%  \centering
  \includegraphics[width=0.32\linewidth]{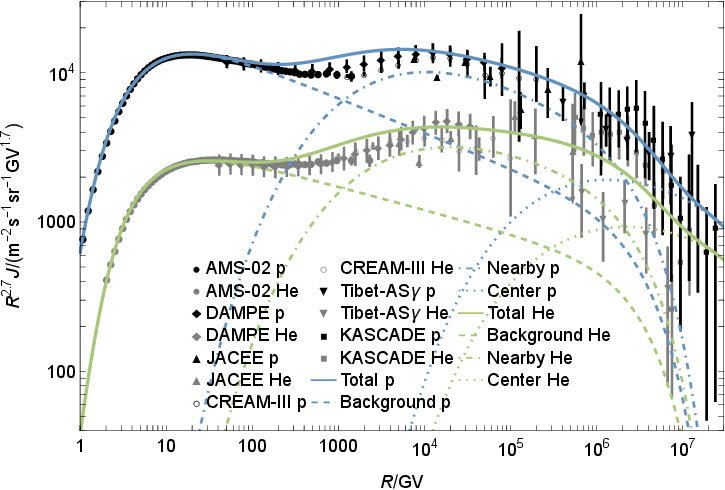}
%\end{minipage}
%\begin{minipage}{0.32\textwidth}
 % \centering
  \includegraphics[width=0.32\linewidth]{spectra.eps}
%\end{minipage}
%\begin{minipage}{0.32\textwidth}
 % \centering
  \includegraphics[width=0.32\linewidth]{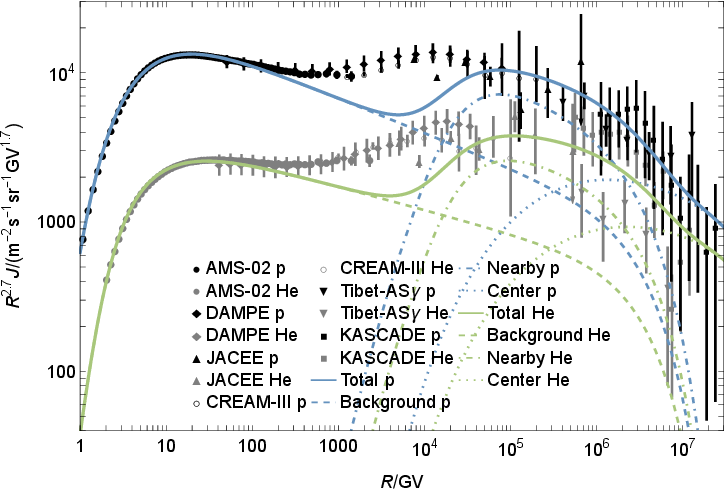}
%\end{minipage}
%\begin{minipage}{0.32\textwidth}
%  \centering
  \includegraphics[width=0.32\linewidth]{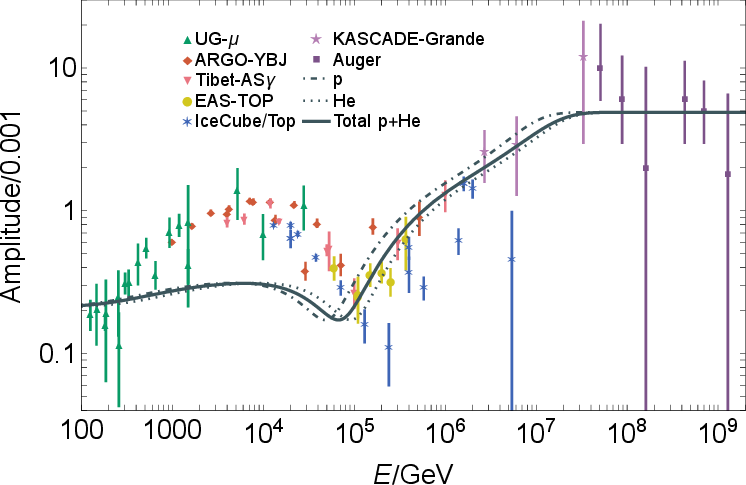}
%\end{minipage}
%\begin{minipage}{0.32\textwidth}
%  \centering
  \includegraphics[width=0.32\linewidth]{anisoamplitude.eps}
%\end{minipage}
%\begin{minipage}{0.32\textwidth}
%  \centering
  \includegraphics[width=0.32\linewidth]{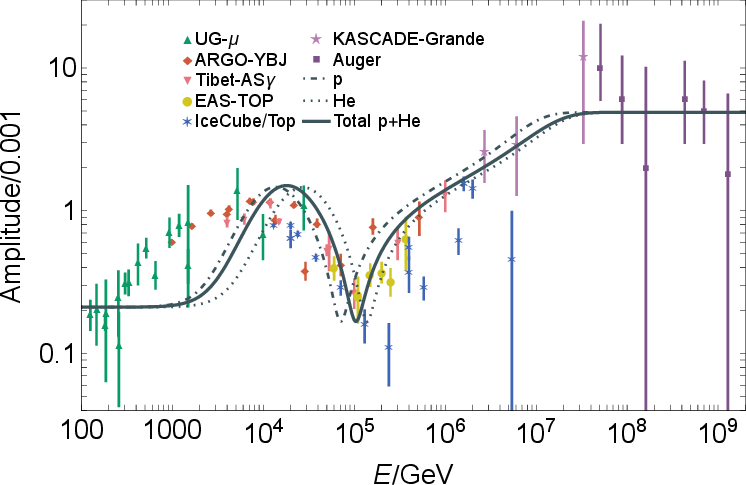}
%\end{minipage}
%\begin{minipage}{0.32\textwidth}
%  \centering
  \includegraphics[width=0.32\linewidth]{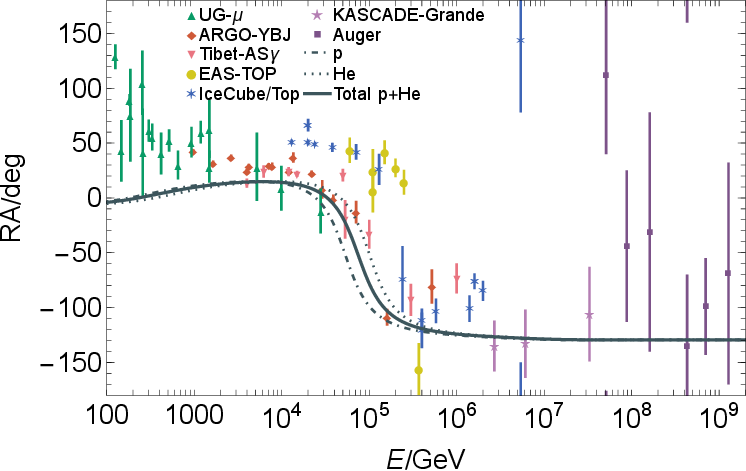}
%\end{minipage}
%\begin{minipage}{0.32\textwidth}
%  \centering
  \includegraphics[width=0.32\linewidth]{anisophase.eps}
%\end{minipage}
%\begin{minipage}{0.32\textwidth}
%  \centering
  \includegraphics[width=0.32\linewidth]{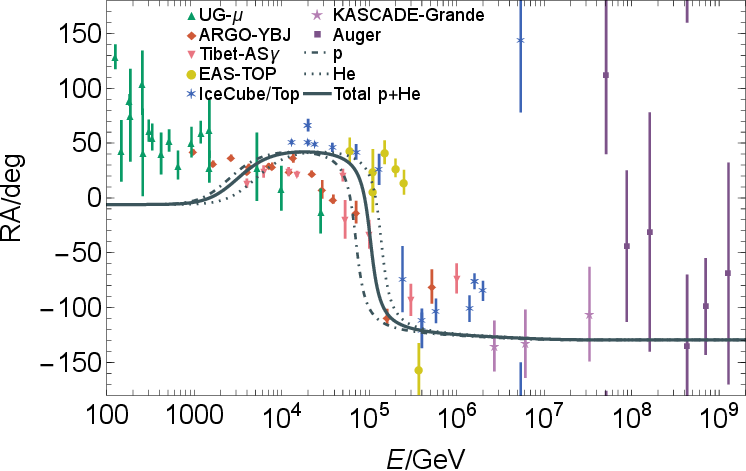}
%\end{minipage}
\caption{Influences of $\rm{V_{\odot}}$ on fitting results. Left column: Fitting results with $\rm{V_{\odot}} = 10$ km s$^{-1}$. Middle column: The best-fitting results with $\rm{V_{\odot}} = 15.5$ km s$^{-1}$. Right column: Fitting results with $\rm{V_{\odot}} = 20$ km s$^{-1}$.}
\label{fig:Vodot}
\end{figure}

Contribution from the nearby source is very sensitive to properties of the local interstellar environment due to CR transport processes in the solar neighborhood. Figure \ref{fig:Vodot} shows both the spectra and dipole anisotropy are very sensitive to solar velocity in the direction of Galactic rotation $\rm{V_{\odot}}$ in the local standard of rest frame. CR observations require a value of about $15.5$ km s$^{-1}$.

\subsection{Dependence on the solar velocity in the LIC  $\bm{{u}_\mathrm{HC}}$}

\begin{figure}[h!]
%\centering
%\begin{minipage}{0.32\textwidth}
%  \centering
  \includegraphics[width=0.32\linewidth]{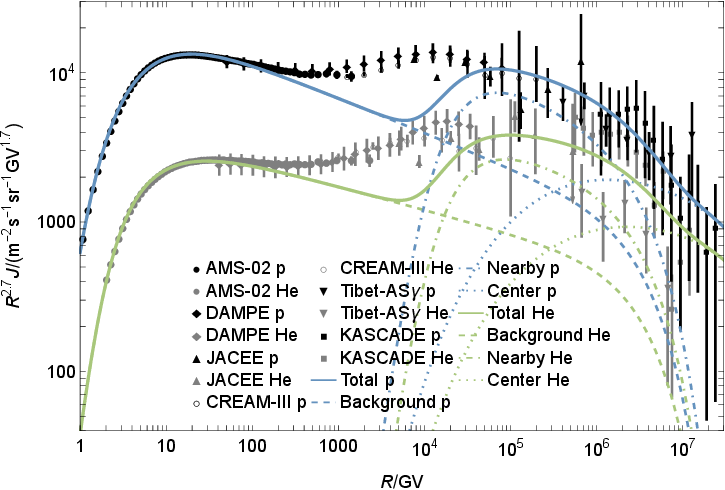}
%\end{minipage}
%\begin{minipage}{0.32\textwidth}
%  \centering
  \includegraphics[width=0.32\linewidth]{spectra.eps}
%\end{minipage}
%\begin{minipage}{0.32\textwidth}
%  \centering
  \includegraphics[width=0.32\linewidth]{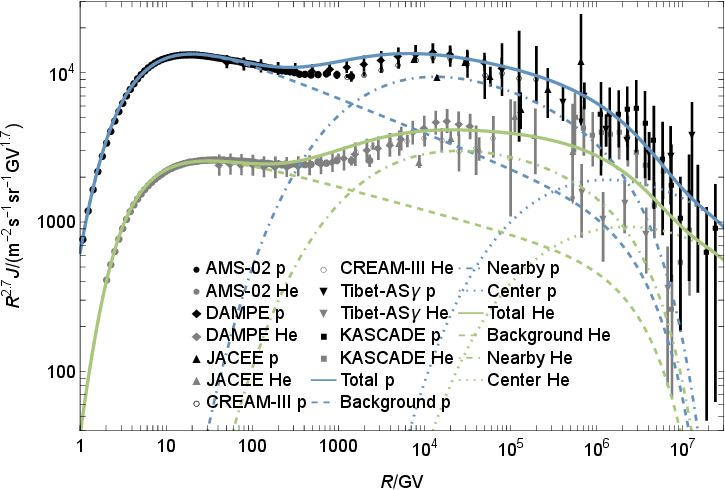}
%\end{minipage}
%\begin{minipage}{0.32\textwidth}
%  \centering
  \includegraphics[width=0.32\linewidth]{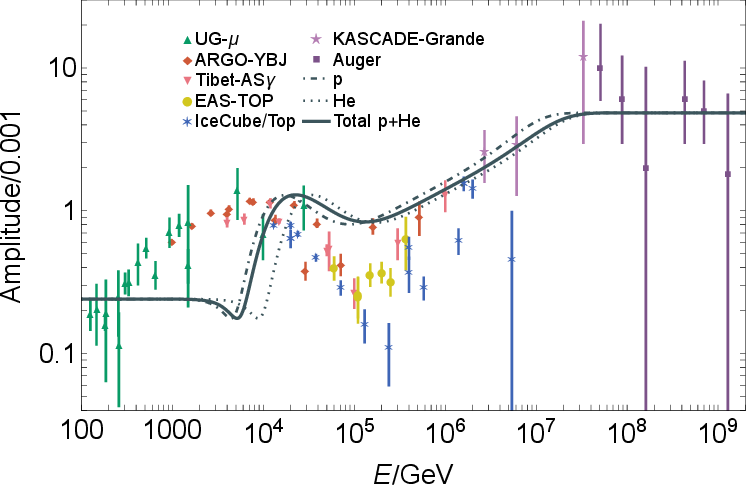}
%\end{minipage}
%\begin{minipage}{0.32\textwidth}
%  \centering
  \includegraphics[width=0.32\linewidth]{anisoamplitude.eps}
%\end{minipage}
%\begin{minipage}{0.32\textwidth}
%  \centering
  \includegraphics[width=0.32\linewidth]{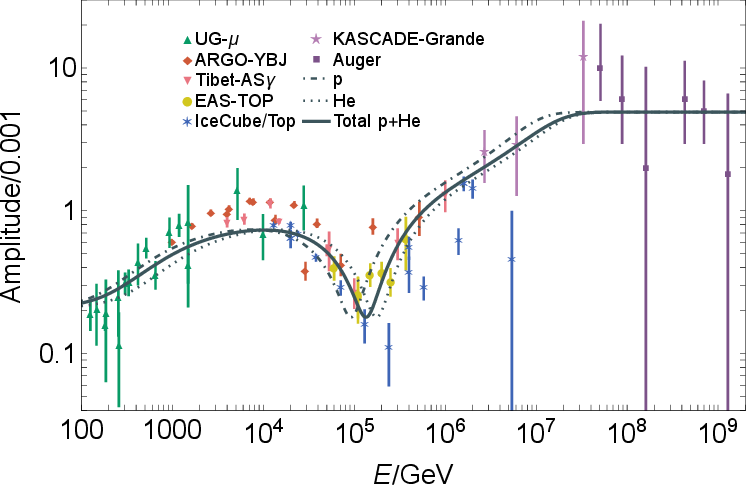}
%\end{minipage}
%\begin{minipage}{0.32\textwidth}
%  \centering
  \includegraphics[width=0.32\linewidth]{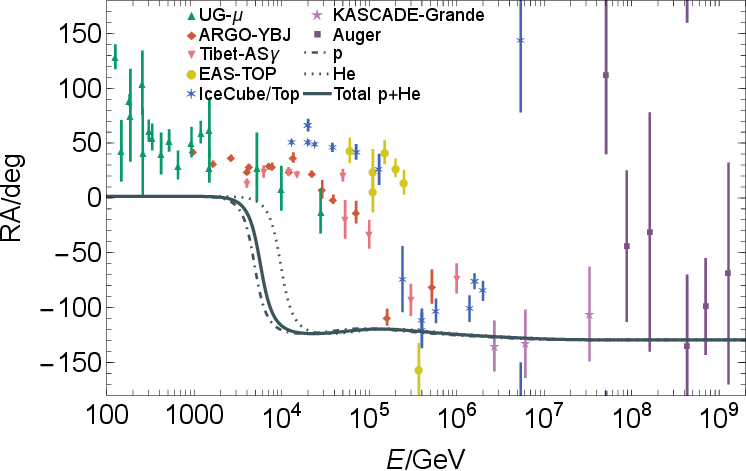}
%\end{minipage}
%\begin{minipage}{0.32\textwidth}
%  \centering
  \includegraphics[width=0.32\linewidth]{anisophase.eps}
%\end{minipage}
%\begin{minipage}{0.32\textwidth}
%  \centering
  \includegraphics[width=0.32\linewidth]{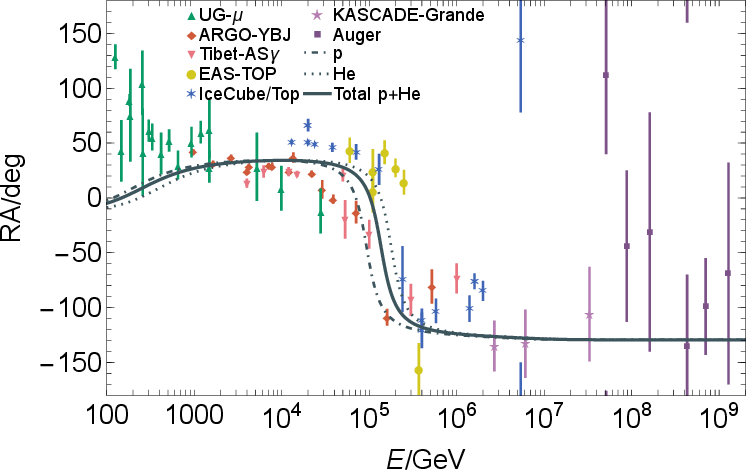}
%\end{minipage}
\caption{Influences of $\bm{{u}_\mathrm{HC}}$ on fitting results. Left column: Fitting results with ${u}_\mathrm{HC} = 23.2$ km/s with a right ascension of 78.5$^\circ$ and a declination of 18.0$^\circ$ \cite{schwadron2014global}. Middle column: The best-fitting results with ${u}_\mathrm{HC} = 25.4$ km/s with a right ascension of 74.9$^\circ$ and a declination of 17.6$^\circ$ \cite{schwadron2015determination}. Right column: Fitting results with ${u}_\mathrm{HC} = 26.6$ km/s with a right ascension of 75.0$^\circ$ and a declination of 17.7$^\circ$ \cite{swaczyna2023interstellar}.}
\label{fig:uHC}
\end{figure}

The velocity of the Sun in the LIC $\bm{{u}_\mathrm{HC}}$ not only affects the Compton-Getting effect, but also the velocity of the LIC in the local standard of rest frame $\bm{{u}_\mathrm{LSR}}$. Figure \ref{fig:uHC} shows that an earlier measurement of $\bm{{u}_\mathrm{LSR}}$ \cite{schwadron2014global} can be ruled out based on our model.

\subsection{Dependence on the characteristic size of the LIC $x$}
\begin{figure}[h!]
%\centering
%\begin{minipage}{0.32\textwidth}
%  \centering
  \includegraphics[width=0.32\linewidth]{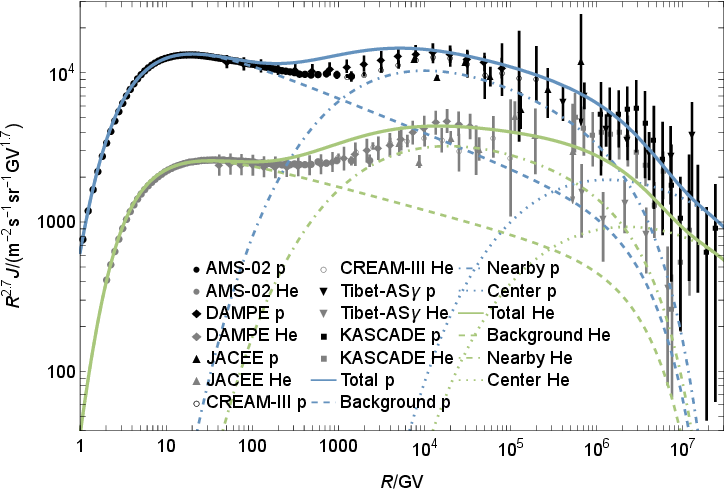}
%\end{minipage}
%\begin{minipage}{0.32\textwidth}
%  \centering
  \includegraphics[width=0.32\linewidth]{spectra.eps}
%\end{minipage}
%\begin{minipage}{0.32\textwidth}
%  \centering
  \includegraphics[width=0.32\linewidth]{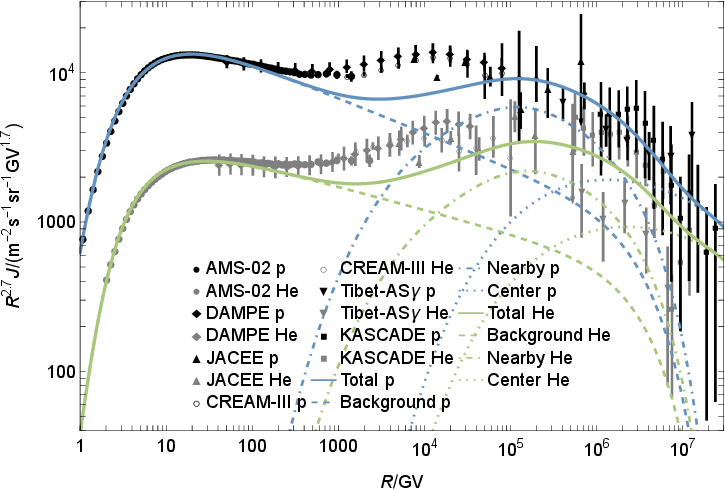}
%\end{minipage}
%\begin{minipage}{0.32\textwidth}
%  \centering
  \includegraphics[width=0.32\linewidth]{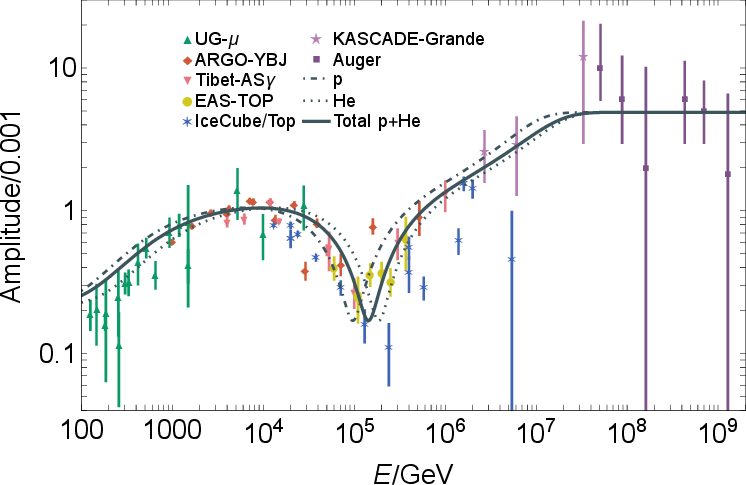}
%\end{minipage}
%\begin{minipage}{0.32\textwidth}
%  \centering
  \includegraphics[width=0.32\linewidth]{anisoamplitude.eps}
%\end{minipage}
%\begin{minipage}{0.32\textwidth}
%  \centering
  \includegraphics[width=0.32\linewidth]{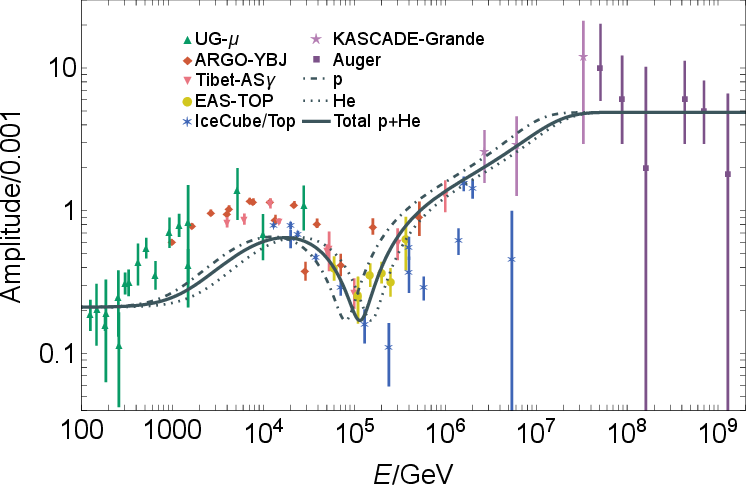}
%\end{minipage
%\begin{minipage}{0.32\textwidth}
%  \centering
  \includegraphics[width=0.32\linewidth]{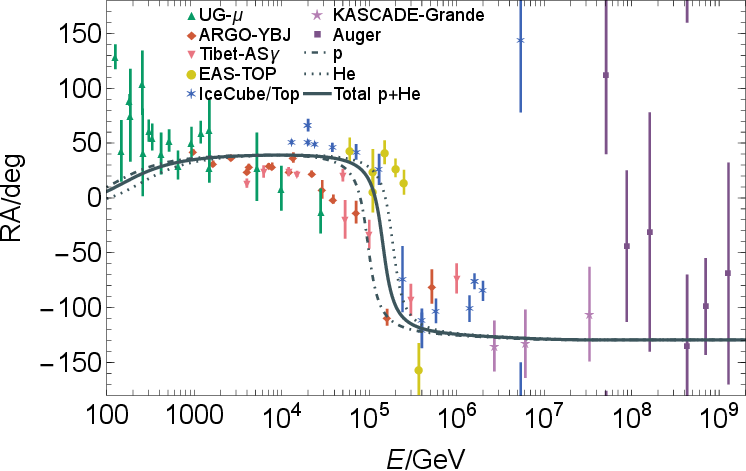}
%\end{minipage}
%\begin{minipage}{0.32\textwidth}
%  \centering
  \includegraphics[width=0.32\linewidth]{anisophase.eps}
%\end{minipage}
%\begin{minipage}{0.32\textwidth}
%  \centering 
\includegraphics[width=0.32\linewidth]{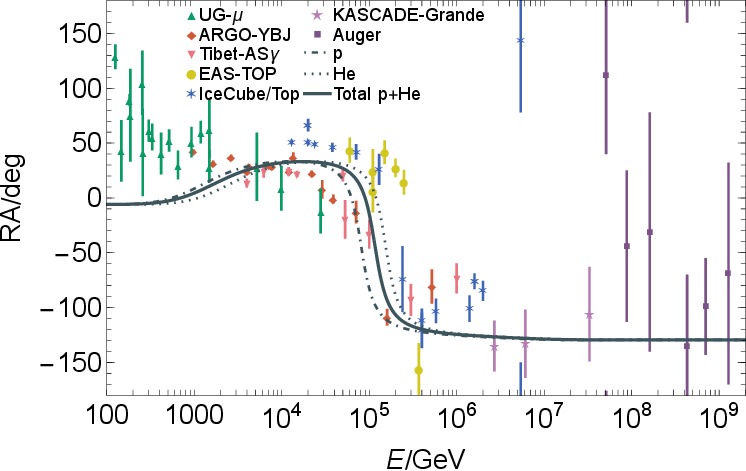}
%\end{minipage}
\caption{Influences of the characteristic size $x$ of the LIC on fitting results. Left column: Fitting results with $x=0.1\;\rm{pc}$. Middle column: The best-fitting results with $x=1.6\;\rm{pc}$. Right column: Fitting results with $x=10\;\rm{pc}$.}
\label{fig:LICx}
\end{figure}

The characteristic size of the LIC $x$ also affects fluxes of CRs from the nearby source. Figure \ref{fig:LICx} shows the dependence of CR properties on $x$. Since a high value of $x$ will suppress contribution to CRs from the nearby source, an upper limit of a few pc can be derived for $x$, which is consistent with the observed size of the LIC.
\subsection{Dependence on the critical rigidity $R_{\rm cr}$ and the magnetic field $B$}

\begin{figure}[h!]
%\centering
%\begin{minipage}{0.32\textwidth}
%  \centering
  \includegraphics[width=0.32\linewidth]{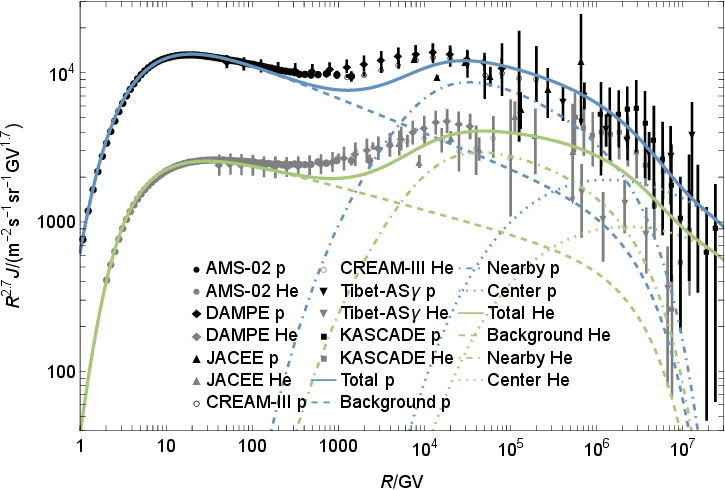}
%\end{minipage}
%\begin{minipage}{0.32\textwidth}
%  \centering
  \includegraphics[width=0.32\linewidth]{spectra.eps}
%\end{minipage}
%\begin{minipage}{0.32\textwidth}
%  \centering
  \includegraphics[width=0.32\linewidth]{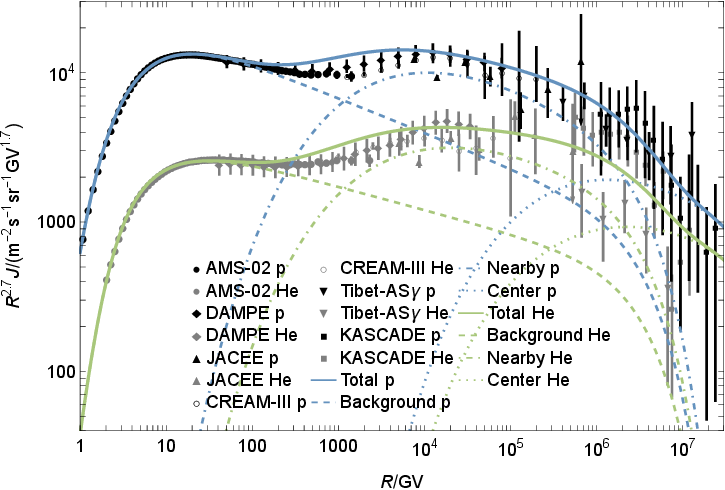}
%\end{minipage}
%\begin{minipage}{0.32\textwidth}
%  \centering
  \includegraphics[width=0.32\linewidth]{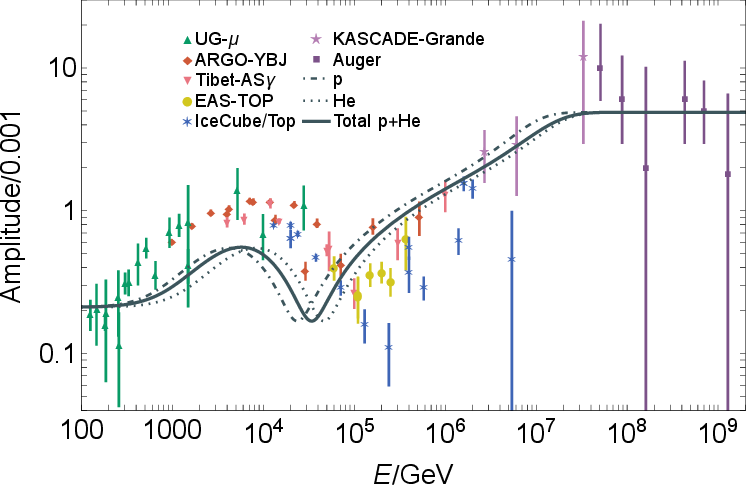}
%\end{minipage}
%\begin{minipage}{0.32\textwidth}
%  \centering
  \includegraphics[width=0.32\linewidth]{anisoamplitude.eps}
%\end{minipage}
%\begin{minipage}{0.32\textwidth}
%  \centering
  \includegraphics[width=0.32\linewidth]{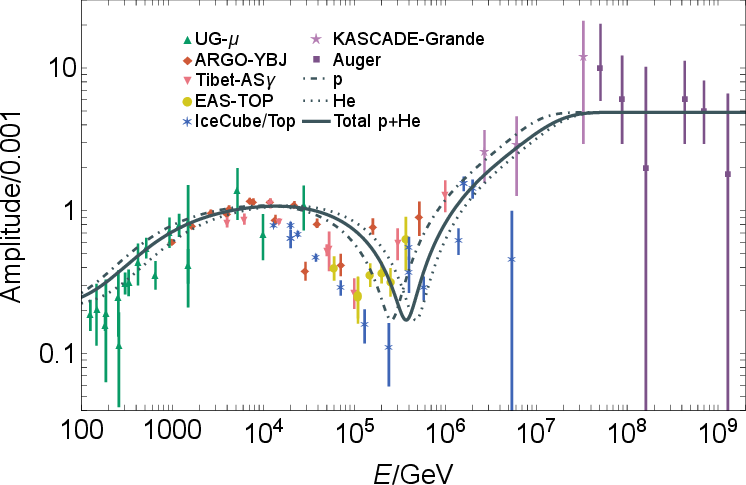}
%\end{minipage}
%\begin{minipage}{0.32\textwidth}
%  \centering
  \includegraphics[width=0.32\linewidth]{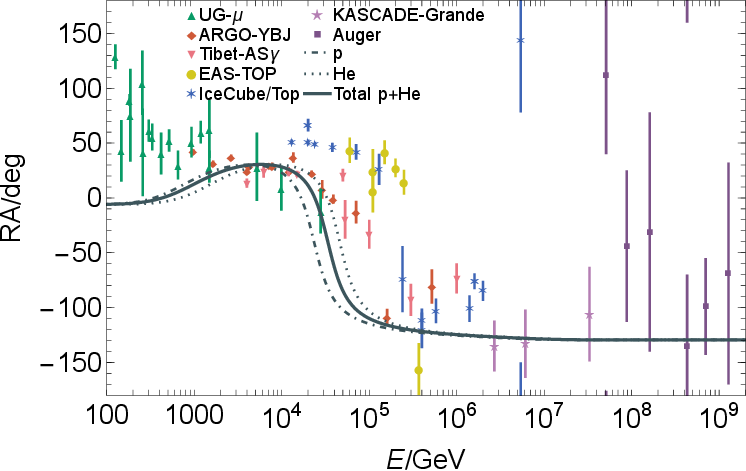}
%\end{minipage}
%\begin{minipage}{0.32\textwidth}
%  \centering
  \includegraphics[width=0.32\linewidth]{anisophase.eps}
%\end{minipage}
%\begin{minipage}{0.32\textwidth}
%  \centering
  \includegraphics[width=0.32\linewidth]{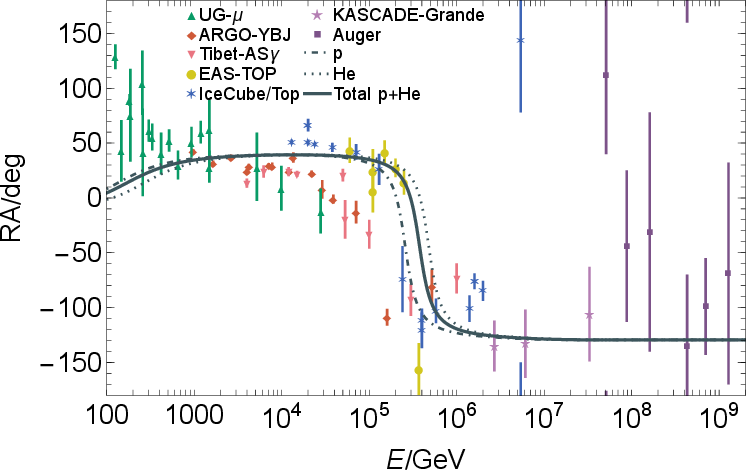}
%\end{minipage}
\caption{Influences of $R_{\rm cr}$ on fitting results. Left column: Fitting results of $R_{\rm cr} = 0.3$ PV. Middle column: The best-fitting results where $R_{\rm cr} = 3$ PV. Right column: Fitting results of $R_{\rm cr} = 30$ PV.}
\label{fig:Rcr}
\end{figure}

We introduce the critical rigidity $R_{\rm cr}$, where Bohm diffusion is achieved, to characterize the level of turbulence in the LIC. For a given magnetic field $B$, a lower value of 
$R_{\rm cr}$ implies a smaller diffusion coefficient $k_{\parallel}$ and a higher turbulence intensity. Figure \ref{fig:Rcr} shows that turbulence in the LIC cannot be very high.

\begin{figure}[h!]
%\centering
%\begin{minipage}{0.32\textwidth}
%  \centering
  \includegraphics[width=0.32\linewidth]{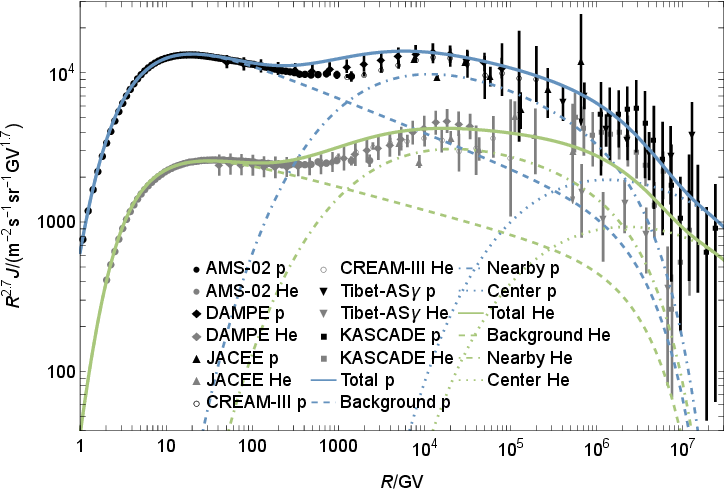}
%\end{minipage}
%\begin{minipage}{0.32\textwidth}
%  \centering
  \includegraphics[width=0.32\linewidth]{spectra.eps}
%\end{minipage}
%\begin{minipage}{0.0.3232\textwidth}
%  \centering
  \includegraphics[width=0.32\linewidth]{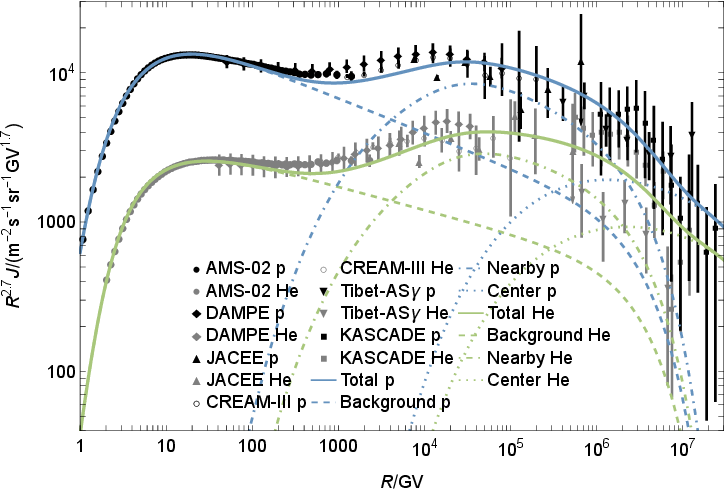}
%\end{minipage}
%\begin{minipage}{0.32\textwidth}
%  \centering
  \includegraphics[width=0.32\linewidth]{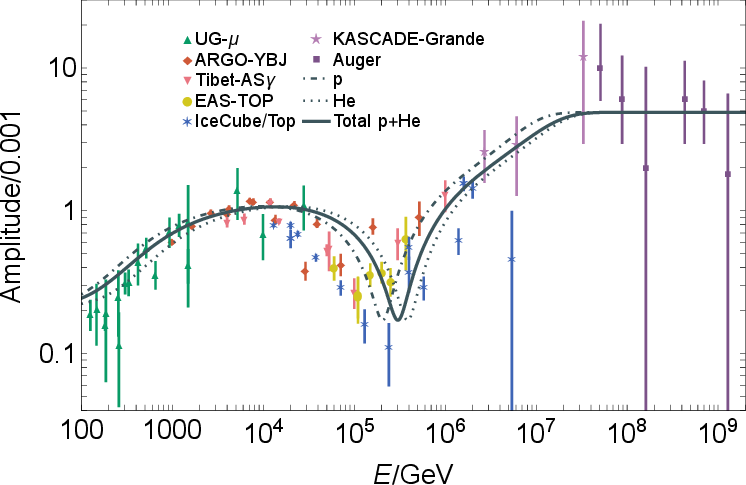}
%\end{minipage}
%\begin{minipage}{0.32\textwidth}
%  \centering
  \includegraphics[width=0.32\linewidth]{anisoamplitude.eps}
%\end{minipage}
%\begin{minipage}{0.32\textwidth}
%  \centering
  \includegraphics[width=0.32\linewidth]{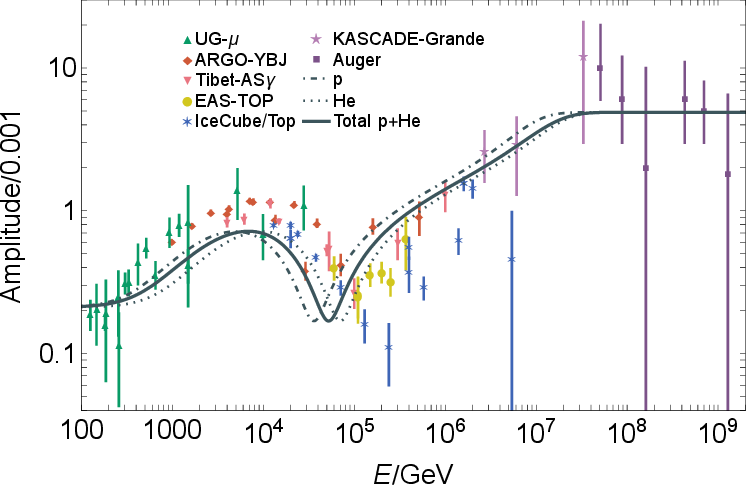}
%\end{minipage}
%\begin{minipage}{0.32\textwidth}
%  \centering
  \includegraphics[width=0.32\linewidth]{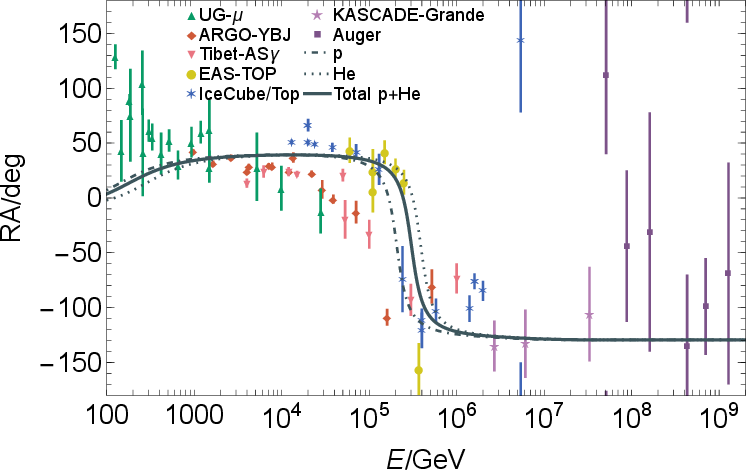}
%\end{minipage}0.32
%\begin{minipage}{0.32\textwidth}
%  \centering
  \includegraphics[width=0.32\linewidth]{anisophase.eps}
%\end{minipage}
%\begin{minipage}{0.32\textwidth}
%  \centering
  \includegraphics[width=0.32\linewidth]{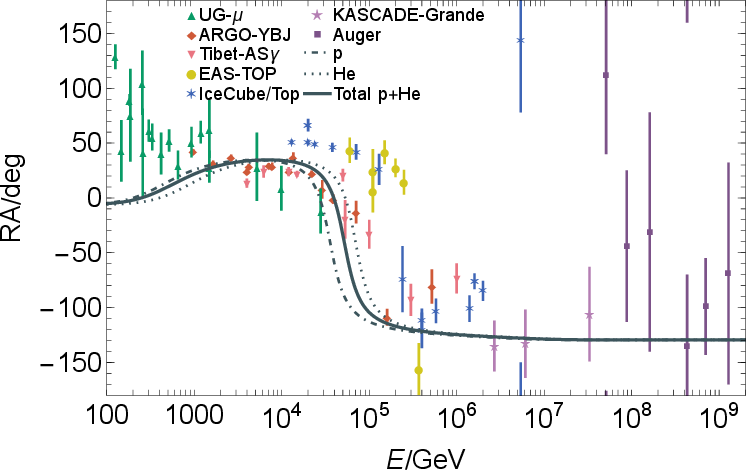}
%\end{minipage}
\caption{Influences of the magnetic field strength $B$ on fitting results. Left column: Fitting results with a magnetic field strength of 1 $\rm{\mu G}$. Middle column: The best-fitting results with a magnetic field strength of 3 $\rm{\mu G}$. Right column: Fitting results with a magnetic field strength of 9 $\rm{\mu G}$.}
\label{fig:B}
\end{figure}

Similarly, for a given $R_{\rm cr}$, $B$ can be used to characterize the turbulence in the LIC. A higher value of $B$ implies a smaller diffusion coefficient $k_{\parallel}$ and a higher turbulence intensity. Figure \ref{fig:B} shows that $B$ is well constrained at a few $\mu$G.

%\subsection{Dependence on the magnetic field strength $B$}

\section{Fitting of KASCADE proton and helium spectra obtained with QGSJet 01 and its influence on the CR dipole anisotropy}
\label{app:spectradata2}

\begin{figure}[h!]
%\centering
%\begin{minipage}{0.32\textwidth}
%  \centering
  \includegraphics[width=0.32\linewidth]{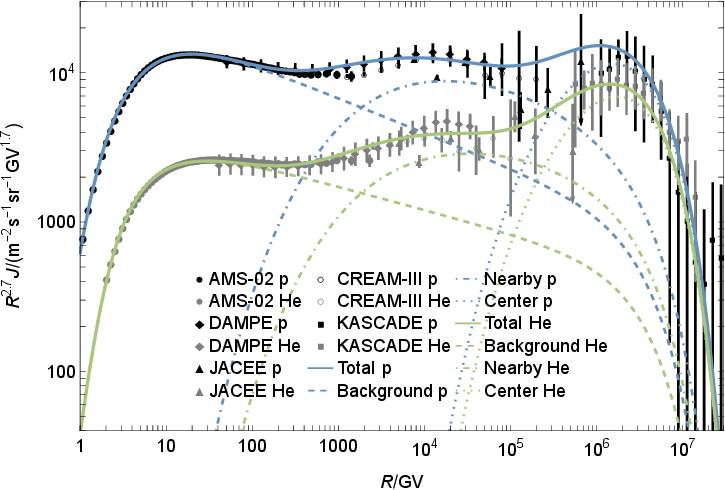}
%\end{minipage}
%\begin{minipage}{0.32\textwidth}
%  \centering
  \includegraphics[width=0.32\linewidth]{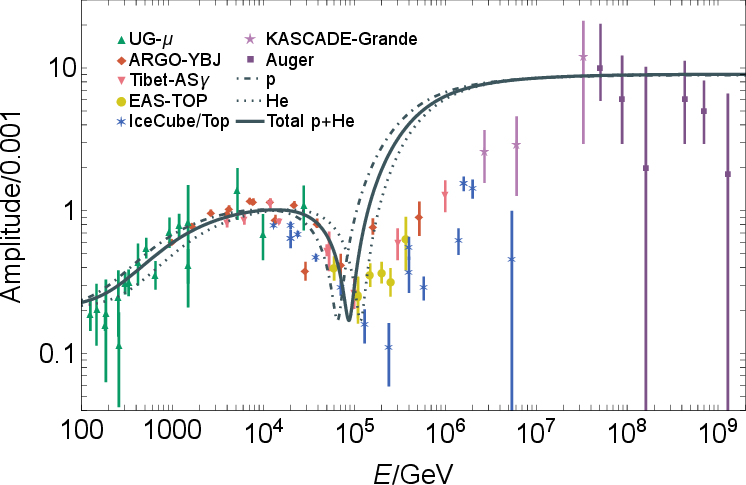}
%\end{minipage}
%\begin{minipage}{0.32\textwidth}
%  \centering
  \includegraphics[width=0.32\linewidth]{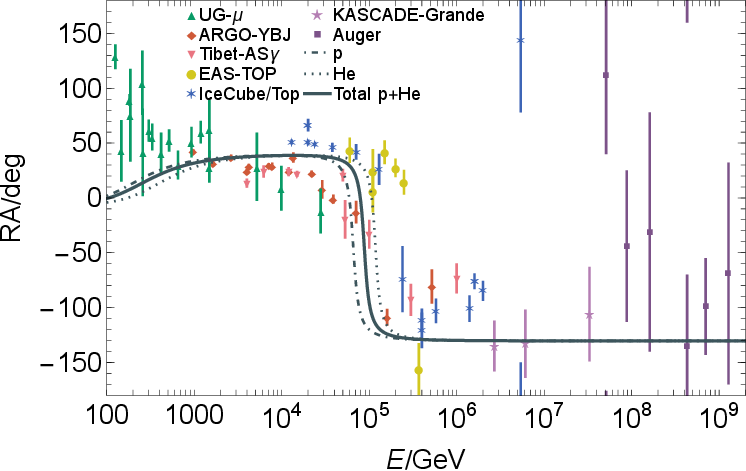}
%\end{minipage}
\caption{The fitting of KASCADE proton and helium spectra obtained with QGSJet 01 and its influence on the CR dipole anisotropy.}
\label{fig:QGSJet}
\end{figure}

To fit the KASCADE proton and helium spectra obtained with QGSJet 01 \cite{antoni2005kascade}, which is about 2 times higher than those obtained with SIBYLL 2.1, we use the same parameters as in the Section~\ref{sec:model} of the main text except for $T_{\rm c}= 2\;\rm{Myr}$, $Q_{\rm 0c,p}=5\times10^{58}\;\rm GV^{-1}$, $Q_{\rm 0c,He}=1\times10^{58}\;\rm GV^{-1}$ and $R_{\rm cutc}=4\;\rm PV$. The fitting results are shown in Fig.~\ref{fig:QGSJet}. In order to fit the amplitude of CR dipole anisotropy above 100 TeV better here, heavier elements need to be considered, which is beyond the scope of the current study.

\bibliography{apstemplate.bib}

\end{document}